\documentstyle[aps,twocolumn]{revtex}
\begin{document}
\draft
\title{ An interpretation of Tsallis statistics based on polydispersity} 
\author{Ramandeep S. Johal\thanks{e-mail: 
raman\%phys@puniv.chd.nic.in}} 
\address{{\it Department of Physics, Panjab University,}\\ 
{\it Chandigarh -160 014, India. }} 
\date{\today}
\def\be{\begin{equation}}
\def\ee{\end{equation}}
\def\ba{\begin{eqnarray}}
\def\ea{\end{eqnarray}}
\maketitle
\begin{abstract}
It is argued that polydispersed sytems like colloids provide a direct 
example where Tsallis' statistical distribution is 
useful for describig  the heirarchical nature of the system 
based on particle size. 
\end{abstract}
\pacs{ 05.20-y, 05.70-a, 61.25.Hq, 64.10.+h}
It is believed \cite{1} that systems which have heirarchical organization 
in phase-space,  involve long-range  interactions, display
spatio-temporal  complexity or have (multi)fractal boundary
conditions, are  not adequately treated within the statistical
framework of Boltzmann and Gibbs. Such sytems are said to 
exhibit nonextensive behaviour and Tsallis has advanced \cite{2} 
a thermostatistical approach based on a generalization
of Shannon entropy as
\be
S^{T}_{q} = \frac{1-\sum_{i=1}^{W} (p_i)^q}{q-1}.
\label{en1}
\ee
Under the generalized constraints \cite{3}, the maximum entropy
principle  yields the following equilibrium distribution
\be
p_i \sim\{1-(1-q){\beta}\varepsilon_i\}^{1/(1-q)},
\label{en2}
\ee
where $\beta$ is the Lagrange multiplier, $\epsilon_i$
is a random variable like energy, satisfying some
constraint analogous to mean value. For $q\to 1$,
we recover Boltzmann or canonical distribution.
Tsallis formalism is finding a growing number of applications 
\cite{4} which do lend support to the validity of underlying postulates,
still it is not fully established as to what physical
principles are involved therein. It has been conjectured \cite{5}
that Tsallis formalism alludes to a scale invariant statistical
mechanics. More precisely, Tsallis-like distribution can be derived
by assuming an ensemble which has replicas of the same system
at different scales. Typical examples of these are the
polydispersed systems like polymers, commercial
surfactants, colloidal suspensions and critical spin
systems on heirarchical lattices. In this letter,
we present an interpretation of  Tsallis distribution
in terms of colloidal polydispersity. 

Recently \cite{6}, for the case of Levy flights, 
the exponential distribution ${\rm exp}(-x/\lambda)$
was mapped onto Tsallis distribution by considering fluctuations
in the parameter $\lambda$ about the mean value $\lambda_0$.
The appropriate distribution function turns out to be Gamma distribution
\be
f\left(\frac{1}{\lambda}\right) = \frac{1}{\Gamma(\alpha)} (\alpha\lambda_0)
                      ^{\alpha} \left(\frac{1}{\lambda}\right)^{\alpha -1}
                      {\rm exp}\left( -\alpha\frac{\lambda_0}{\lambda}
                      \right).
\label{en3}
\ee
Thus 
\be
\int_{0}^{\infty} e^{-x/\lambda}\; f\left(\frac{1}{\lambda}\right) 
\;{d}\left(\frac{1}{\lambda}\right)=
\left(1+ \frac{1}{\alpha}\frac{x}{\lambda_0}\right)^{-\alpha},
\label{en4}
\ee
which on identifying $\alpha = 1/(q-1)$, becomes Tsallis-like 
distribution, Eq. (\ref{en2}). The Tsallis index is given in
terms of the relative variance 
\be
\omega = \frac{ \langle\left(\frac{1}{\lambda}\right)^2\rangle -
         \langle \frac{1}{\lambda}\rangle^2}{\langle 
         \frac{1}{\lambda}\rangle^2} = \frac{1}{\alpha} = (q-1).
\label{en5}
\ee
Now thermodynamically, a   polydispersed colloid 
can be treated as a system with continuously infinite
species type \cite{war1}-\cite{war4}. The fluctuating
parameter in this case may be taken as the particle size $\sigma$.
Moreover, it is usual to describe particle sizes by Gamma
or Schulz distribution \cite{shul}
\be
p(\sigma)  = \frac{1}{\gamma^{\alpha}\Gamma(\alpha)}
             \sigma^{\alpha -1}\; e^{-\sigma/\gamma},  
\label{en6}
\ee
where we have put $\gamma = \frac{\bar{\sigma}}{\alpha}$ and 
$\bar{\sigma} = \int \sigma p(\sigma)d\sigma$.
Again using $\alpha = 1/(q-1)$, we find that Tsallis
distribution over variable $\theta$ can be defined as
\be
(1-(1-q)\bar{\sigma}\theta)^{1/(1-q)} = \int e^{-\sigma\theta}
                                       \;p(\sigma)\;d\sigma.
\label{en7}
\ee
In a different language \cite{ellis}, r.h.s. of the above equation defines
the cumulant generating function $h(\theta)$ as
\be
h(\theta) = {\rm ln}\int e^{-\sigma\theta} \;p(\sigma)\;d\sigma.
\label{en8}
\ee
In other words, cumulant generating function for Schulz distribution
is the natural logarithm of Tsallis distribution, (Eq.(\ref{en7})). 

Futhermore as established in \cite{war98}, $h(\theta)$ is related
to combinatorial entropy per particle (which appears
in the joint free energy on mixing two systems, for details
see \cite{war98})  by a Legendre transform
\be
{\rm ln}\;P(m)/N = h(\theta) + m\theta,
\label{en9}  
\ee
where  $N$ is the number of particles in one  
subsystem and $m =- \frac{\partial h}{\partial \theta}$ is the moment 
variable which was taken to be mean size $\sum_i \sigma_i/N$. 

From Eq. (\ref{en9}), we see that there is a direct relation
between Tsallis distribution and combinatorial entropy.
From standard thermodynamics, we know that entropy $S$
and free energy $\Psi$ are related by Legendre transform
$S(M) = -\Psi(\beta) + \beta M$, 
where $\beta$ is called intensity and mean value $M$ is called
extensity. Thus noting that $\theta\equiv\beta$, $M\equiv
m$, we infer that $-h(\theta)$ is a kind of generalized free
energy per particle. Also as free energy
$\Psi = -{\rm ln}\;Z$, where $Z$ is the partition function,
so from Eq. (\ref{en8}) we conclude that Eq. (\ref{en7})
represents generalized partition function per particle. 

We see from Eq. (\ref{en5}) that relative variance of particle
size with respect to Schulz distribution is equal to
$(q-1)$. In other words, the degree of nonextensivity
$(q-1)$ is directly reflected in the polydispersity.
As the relative variance goes to zero (or $q\to 1$), we have 
monodisperse colloidal system and we expect that statistical
description of the system goes from Tsallis to 
Boltzmann distribution.  

Note that  it is not implied that statistical mechanics
of polydispersity is completely equivalent to Tsallis statistics. 
Rather we  propose that heirarchical nature of systems 
describable under Tsallis formalism is captured by polydispersity.
Tsallis formalism also treats systems with long range interactions
and it doesn't seem correct to say that such interactions are "switched on" 
in going from mono- to polydisperse system e.g. in charged stabilised
colloidal particles interacting through screened Coulomb potential. 
On the other
side,  statistical mechanics of 
polydispersity has been studied using different approaches 
\cite{war98}-\cite{lado}. 
The connection with
Tsallis formalism can be hoped to contribute towards these approaches by 
incorporating the self similar (heirarchical) nature of the 
colloidal systems.

Finally, we make a remark about alternative approaches for
mapping exponential distribution to Tsallis distribution.
This has been done using Hilhorst integral transformation \cite{curl}.
In a recent work \cite{jo258}, the equivalence of Tsallis distribution
with a modification of the exponential distribtion, based
on quantum groups was studied. The latter distribution
in the present case  is given as $q_g^{-\bar{\sigma}\theta/(q_g-1)}$,
where $q_g>1$ is the deformation or quantum group parameter. 
(For $q_g\to 1$, we recover $e^{-\bar{\sigma}\theta}$.)
This function can be written as $e^{-{\sigma}\theta}$,
where $\sigma = \bar{\sigma}\frac{{\rm ln}\;q_g}{q_g-1}\equiv 
\bar{\sigma}u$. Thus fluctuations in parameter $\sigma$ as
employed above, can effectively be considered as fluctuations in
$u$. In terms of $u$, then one can write Tsallis distribution as
\be
\left(1+\frac{\bar{\sigma}\theta}{\alpha}\right)^{-\alpha} = 
\int e^{-\sigma\theta u} \;p(u)\;du,
\label{10}
\ee   
where as before $\alpha = 1/(q-1)$, and 
\be
p(u) =  \frac{\alpha^{\alpha}}{\Gamma(\alpha)}
             u^{\alpha -1}\; e^{-u\alpha}. 
\label{11}
\ee
Also note that $\bar{u} = \int up(u)\;du = 1$. The above
distribution actually defines Mellin's Transformation, which
has been used for a similar purpose in the context of
thermal field theory \cite{abe}.  

\end{document}